\documentclass[aps,prb,twocolumn,groupedaddress]{revtex4}
\usepackage{makeidx}
\usepackage{amsmath,amssymb,amsfonts,amsthm}
\usepackage{graphicx,bm}

\begin{document}

\title{On the recently proposed martensitic-like structural transformation
in V, Nb, and Ta}

\author{F. Cordero}
\address{CNR-ISC, Istituto dei Sistemi Complessi, Area della Ricerca
di Roma - Tor Vergata, Via del Fosso del Cavaliere 100, I-00133
Roma, Italy}

\begin{abstract}
A recent announcement, after anomalies in the thermal expansion,
of a previously unknown martensitic transformation in pure Nb, Ta and V
is discussed in the light of anelastic measurements on Nb samples with
comparable purities.
It results that the effects of the alleged transformation on the
anelastic spectra would be at least three orders of magnitude
smaller than in typical martensitic transformations. A possible
alternative explanation for the observed anomalies is proposed
in terms of precipitation of unnoticed
residual H.
\end{abstract}


\maketitle

\section{Introduction}
In an intriguing Letter \cite{BWN11} it has been reported that
crystals of
pure Nb, Ta and V exhibit a splitting of the thermal expansion along the $%
\left\langle 100\right\rangle $ directions below room temperature,
attributed to a previously unknown martensitic transformation (MT). The
occurrence of the MT would be prevented by impurities at levels as low as
few hundreds at ppm, so explaining why it had never been noticed before \cite%
{BWN11}. It is even postulated that the structural instability in
A-15 intermetallics like Nb$_3$Sn is due to the presence of the
Group-Va element itself \cite{BWN11}, rather than to their
correlation in one-dimensional chains, as usually accepted
\cite{Tes75}. This is a provocative finding, since Nb, Ta and V have
been thoroughly studied for decades starting from the 1960s, and are
known to have $bcc$ structure at all temperatures. In order to
induce a structural instability in V,\ a pressure of 69~GPa must be
applied, so causing a rhombohedral distortion of the cell
\cite{DAS07}, while Nb exhibits some shallow softening of the
$c_{44}$ elastic constant above room temperature \cite{TWS77}, and
Ta presents the same effect at $\sim 80$~GPa \cite{AFS10}, but
neither Nb nor Ta have been found to complete the transition to the
rhombohedral state.

Many investigations on these group Va transition metals aimed at
studying dislocations and their interaction with the gaseous
impurities H, O, N and C, while object of other investigations were
the absorption, diffusion and trapping of such impurities,
particularly H, and the complex $x-T$ phase diagrams of MH$_{x}$ (M
= V, Nb, Ta) \cite{AV78,Fuk93,Wip97}. Many of these investigations
have been conducted by anelastic relaxation and neutron scattering,
which are also the most sensitive in detecting martensitic
transformations, but none had been detected so far.

Also the electronic structure of these transition metals has been thoroughly
studied both experimentally and theoretically and has been shown to be at
the origin of the tendency or manifestation of structural instability at
high temperature and pressure. The softening of the shear elastic constant $%
c_{44}$ is driven by a combination of intraband nesting of the Fermi
surface, electronic topological transition, and band Jahn--Teller
effect \cite{AFS10,LKS06b}. When the softening becomes complete, as
in V at high pressure, the lattice becomes unstable against shears
of the $\varepsilon _{4}$ type (Voigt notation) and becomes
rhombohedral, but in Nb and especially Ta this instability is more
than counterbalanced by the Madelung contribution to the elastic
energy \cite{AFS10,LKS06b,WB80}. Tantalum is even considered as
"prototype metal for the investigation and calibration of equation
of state and material strength at extreme thermodynamics conditions"
\cite{AFS10}.

For these reasons, a report of martensitic-like transformations so
far unnoticed in V, Nb and Ta should be carefully considered and
verified also by the methods most sensitive in detecting MTs.
Experimental techniques widely used to study the MTs are diffraction
experiments, revealing splittings of the Bragg peaks, TEM revealing
the twin domains of the low symmetry phase, and anelastic or
ultrasonic experiments exhibiting cusped or steplike softening of
the shear moduli involved in the transformation and elastic energy
loss from the movement of the twin boundaries. None of these
features have ever been reported in pure and dislocation-free V, Nb
and Ta. Yet, the lack of evidence of MT from the existing studies
might be due to the fact they were mostly made on samples with
contents of impurities, either added or unwanted or unknown, that
would hinder the transformation. Indeed, anelastic relaxation
experiments have been done also on crystals much purer than those of
Ref. \cite{BWN11}, for example on a Ta crystal with residual
resistivity ratio RRR = 17000 measured with the torsion pendulum
\cite{BBS89}, and did not show hints to phase transformations. These
experiments, however, were devoted to studying dislocations on
deformed samples, so that the reported anelastic spectra are
dominated by the motions of dislocations. The search for a MT
hindered by minimal amounts of impurities was not an issue and to my
knowledge there are no reports of the background complex elastic
moduli of undeformed samples with very high purity. Since the level
of impurities seems to be critical in revealing the MT, here are
presented the anelastic spectra of two samples of
Nb with purities as close as possible to those where a MT appears in Ref. %
\cite{BWN11}.

\section{Experimental}

Sample \#1 was a $50\times 5\times 0.67$~mm$^{3}$ polycrystalline
bar prepared by Prof. G. H\"{o}rz (MPI f\"{u}r Metallforschung,
Inst. f\"{u}r Werkstoffwissenschaften, Stuttgart). The residual
resistivity ratio was RRR~= $R\left( 296~\text{K}\right) /R\left(
0~\text{K}\right) =$ 320, very close to the value of 347 of the pure
Nb crystal of Ref. \cite{BWN11}, from which a residual resistivity
$\rho \left( 0~\text{K}\right) =6.13\times 10^{-2}$~$\mu \Omega $cm
is deduced. Assuming that the main contribution is from interstitial
O, which contributes with $4.5\times 10^{-10}$~$\Omega
\times $cm/at ppm O \cite{SFS76}, the impurity content was $c_{\text{O}%
}\lesssim 136$~at ppm, the\ $<$ sign being due to the fact that the
anelastic spectrum showed also the presence of H. Sample \#2 was a
$40\times 4.5\times 0.56$~mm$^{3}$ bar cut from Marz grade
polycrystalline Nb, previously subjected to various thermal
treatments during which O uptake occurred. The O content estimated
from RRR = 26.5 was $c_{\text{O}}\simeq $ 1200 ppm, about ten times
larger than in sample \#1, but close to that of the pure Ta crystal
of Ref. \cite{BWN11}. The impurity content of sample \#2 might also
be higher than those of the heat treated samples of Ref.
\cite{BWN11}, which are not specified. The dynamic Young's modulus
$E\left( \omega ,T\right) =E^{\prime }-iE^{\prime \prime }$ was
measured by suspending the samples on thin thermocouple wires and
electrostatically exciting the flexural modes \cite{135}; the first
and fifth flexural modes, whose nodal lines practically coincide and
whose frequencies are in the ratio 1:13.3, could be measured during
a same run.

\section{Results and Discussion}

Figure 1 presents the anelastic spectra of the above samples: the upper
panel contains the relative change of $E^{\prime }$ with respect to its 0~K
value, and the lower panel contains the elastic energy loss coefficient $%
Q^{-1}=$ $E^{\prime \prime }/E^{\prime }$. Of sample \#1 are shown the $%
Q^{-1}\left( T\right) $ curves measured at both 1 and 13~kHz (closed
symbols). The almost linear rise of $Q^{-1}\left( T\right) $ is
thermoelastic effect \cite{NB72,CC76}, namely the diffusion of heat
between the alternately expanded and compressed faces of the bar
during the flexural vibration, with consequent out-of-phase
expansion of the heated region. In sample \#2 (open symbols) there
is no trace of additional dissipation
mechanisms, while in sample \#1 (closed symbols) there are two peaks at $%
\sim 90$~K and $\sim 190$~K. These peaks are shifted to higher $T$
when measured at higher frequency (smaller symbols) and therefore
are due to thermally activated relaxations. The peak at $\sim 90$~K
is readily recognized from its temperature and activation energy as
due to the hopping of residual H trapped by O \cite{MB75b,CC82},
while the peak at $\sim 190$~K is associated with the simultaneous
presence of H and dislocations \cite{BB72,NPP01}. It turns out that,
in spite of the fact that sample \#1 has a RRR 12 times higher than
that of sample \#2, its content of residual H and dislocations is
larger. The seeming inconsistency is due to the fact that H
contributes little to the residual resistivity, especially when it
is trapped by an impurity like O, and dislocations contribute even
less. Therefore, the RRR provides a good estimate of the content of
heavy gaseous impurities, but neither of H nor of dislocations. The
$\sim 130$~at ppm O impurities can trap about as many H atoms, and
if untrapped H in excess is also present, it would partially
precipitate into $\beta $ phase hydride. When this precipitation
from the the gas-like $\alpha $ phase to the $\alpha +\beta $ phase
occurs, it is accompanied by the formation of dislocations at the
phase interfaces and manifests itself in the anelastic spectrum as a
sharp rise of dissipation and anomalies in the modulus
\cite{AV78,Fuk93,2,YK82a}. The absence of a clear precipitation peak
indicates that, if any additional H was present in solid solution,
its precipitation might have only occurred below 150~K, and
therefore could not exceed hundred at ppm. One-two hundred at ppm is
the typical residual content of H even after prolonged outgassing
treatments in UHV, unless particular precautions are taken, like
applying a film of Pd on the sample surface \cite{BBS89}, or an
oxide layer during cooling from the high temperature annealing
\cite{KMO80}.

\begin{figure}[tbp]
\includegraphics[width=8.5 cm]{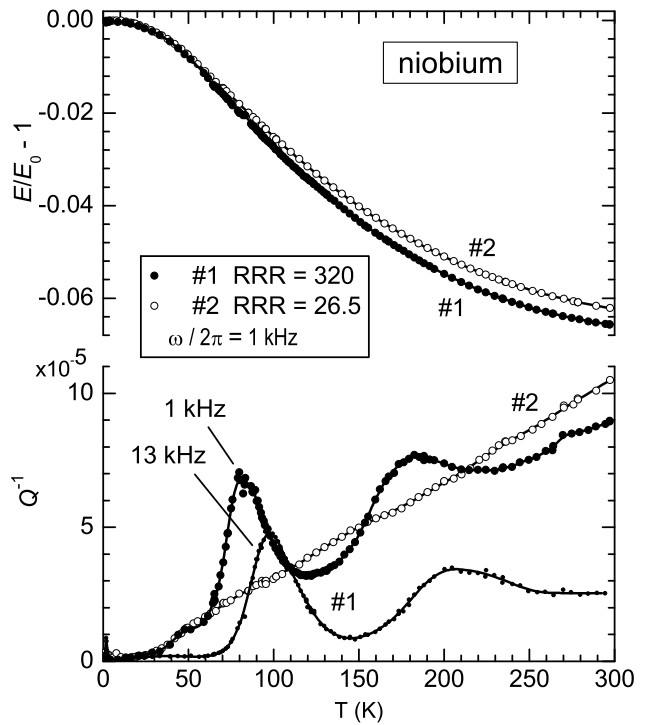}
\caption{Relative change of the Young's modulus $E$ and elastic
energy loss coefficient $Q^{-1}$ $=$ $E^{\prime \prime }/E^{\prime
}$ measured on two samples of Nb with different purities.}
\label{fig1}
\end{figure}

The real parts of the modulus do not present any anomaly attributable to
structural instabilities, except for the well known shallow softening above
room temperature from the Fermi surface topology and electron-phonon
coupling, which appears as a positive curvature of the $E\left( T\right) $
curve. There is a drop of $E\left( T\right) $ below the superconducting
transition at 9.2~K, again due to the electron-phonon coupling, but is
invisible on the scale of Fig. 1.

The curves in Fig. 1 exclude the occurrence of any structural
transformation near or below room temperature with a sensitivity
much higher than diffraction experiments. In fact, a MT would appear
as a steplike or cusped softening \cite{Lut07} with amplitudes up to
several tens of percent in the polycrystalline Young's modulus $E$
\cite{BCM03}, which contains all the elastic constants, but no trace
of it is found in Fig. 1. In addition, the motion of the twin walls
formed in the low symmetry phase would cause a very broad thermally
activated maximum with sharp onset at the transformation,
whose typical amplitude is \cite{Lut07,BCM03} $\Delta Q^{-1}=10^{-3}-10^{-1}$%
. A similar anomaly, if present in Fig. 1, cannot have an amplitude above
few $10^{-6}$. It can be concluded that, if a MT occurs in any of the two
samples, it causes elastic and anelastic anomalies with an amplitude at
least three orders of magnitude smaller than in the known cases of MTs.

It should be stressed that a transformation where the order
parameter is strain (ferroelastic transformation), is best detected
in the complex elastic modulus (or its reciprocal, the compliance),
exactly as a ferroelectric or magnetic transition appears in the
dielectric and magnetic susceptibilities. Therefore, in the absence
of a clear splitting of the Bragg peaks in diffraction experiments,
the hard evidence for a ferroelastic transformation should come from
elastic and/or anelastic anomalies, just like the case of the
ferroelectric and ferromagnetic transitions, where the main evidence
is the Curie-Weiss peak in the respective susceptibility. This is
true also if strain is not the primary order parameter, but only
coupled to it, so causing a possibly small step-like anomaly in the
compliance, rather than a Curie-Weiss peak \cite{Reh73}. In fact,
whatever the nature of the transition, if the cell departs from the
cubic shape, it will lead to the formation of domain walls, whose
motion enhances the mechanical loss to levels much higher than the
$Q^{-1}\left( T\right) $ curves in Fig. 1.

These data show that the absence assumed so far of structural
transformations near and below room temperature in Nb is not due to
a high amount of impurities, since it persists at the same content
of impurities as in Ref. \cite{BWN11}. In addition, the anelastic
spectrum is much more sensitive and selective in characterizing the
status of the sample than resistivity, which is little sensitive to
H and even less to dislocations. It can be concluded that the
interpretation of the thermal expansion anomaly in Nb, Ta and V as
due to a MT \cite{BWN11} is problematic. Lacking an anelastic
characterization of those samples, an alternative explanation can
only be speculated in terms of unwanted interstitial H and
dislocations. In fact, the estimate in Ref. \cite{BWN11} of
$<0.05$~at\% H from the Vickers hardness $H_{V}$ and lattice
parameters \cite{IMU06} seems
unreliable, since the error bars and dispersion of points in the plot of $%
H_{V}$ vs $x$ in Fig. 5 of Ref. \cite{IMU06} (setting $c_{H}\equiv
x$ in NbH$_{x}$) do not allow estimates better than $\Delta x\sim
0.005$. This estimate may still seem to exclude that the
precipitation of H is at the origin of the anomalies with onset just
below room temperature, since the solvus line $x\left( T\right) $
separating $\alpha $ and $\alpha +\beta $ phases in NbH$_{x}$ is
close to 0.03 at room temperature. Yet, the reported solvus line is
not a true border at thermodynamic equilibrium, but presents large
hysteresis between heating and cooling and can be shifted to higher
temperature of tens of kelvins just by repeating the temperature
cycles \cite{Wes76,Fuk93}. This is due to the fact that once the
precipitates of $\beta $ phase are plastically accommodated, the
plastic deformation remains and allows subsequent precipitations to
occur at higher temperature. As a consequence, the anomalies found
in Nb \cite{BWN11} are not incompatible with precipitation of H with
$x\sim 0.01,$ if they are measured after the first cooling run. This
is especially true for the anomalies in V and Ta, which appear well
below room temperature and might be accounted for by a much lower
content of H.

Another issue in the interpretation of the anomalies in the thermal
expansion in terms of MT \cite{BWN11} is their anisotropy on a
macroscopic scale. It is unlikely that, in crystals with edges long
up to 3.5~cm, the transformation strains of the twin domains do not
average in all directions and instead the edges present different
elongations, unless the crystals are strained or rich in
dislocations. Analogously, ferromagnetic or ferroelectric materials
do not acquire macroscopic magnetization or polarization in the
absence of an external field or anisotropic defects.

\section{Conclusion}

In conclusion, a convincing evidence of the existence of a martensitic-like
transformation in Nb, Ta and V can come from anelastic or diffraction
experiments, but the existing anelastic measurements, including those
presented here, provide a negative answer so far.

\end{document}